\documentstyle[12pt]{article}

\newcommand{\eq}[1]{Eq.~(\ref{#1})}
\newcommand{\eqs}[2]{Eqs.~(\ref{#1}) and~(\ref{#2})}

\newcommand{\ra}{\rangle}
\newcommand{\la}{\langle}
\newcommand{\be}{\begin{eqnarray}}
\newcommand{\en}{\end{eqnarray}}

\begin{document}
\author{J.~W.~Bos$^1$, D.~Chang$^{2,1}$, S.~C.~Lee$^1$,
  and H.~H.~Shih$^1$
  \\
  {\small ${}^1$Institute of Physics, Academia Sinica, Taipei, Taiwan}
  \\
  {\small ${}^2$NSC Center for Theoretical Sciences and Department of 
Physics,}
  \\
  {\small ${}^2$ National Tsing Hua University, Hsinchu, Taiwan}
  \\
  }

\title{The Okubo relation for the baryon magnetic moments and chiral 
perturbation theory}

\date{}

\maketitle

\begin{abstract} 
We analyze the Okubo $SU(3)$ relation among the hyperon
  magnetic moments in the usual scheme of chiral perturbation theory
  (ChPT). We classify the one-loop diagrams, including those with 
intermediate decuplet baryons, in a simple way according to
  whether or not they satisfy the Okubo relation.  Contrary to the
  conventional wisdom, we find that one-loop contributions to the hyperon 
magnetic moments in general violate the Okubo relation if the physical
  masses are employed for the meson propagators in the loops.
\end{abstract}
  
Keywords: baryon magnetic moments, chiral perturbation theory, 
  Okubo relation. (Pacs 13.40.Em, 14.20.-c, 11.30.Rd)

\section{Introduction}\label{introduction}

The magnetic moments of the octet baryons were found to obey
approximate $SU(3)$ symmetry a long time ago by Coleman and Glashow
\cite{coleman}.  The relations of Coleman and Glashow are satisfied by
the observed magnetic moments up to about the $20 \%$ level.  Shortly 
thereafter Okubo \cite{okubo} derived a relation between the magnetic
moments, based on the assumption that $SU(3)$ symmetry is broken 
linearly, which is satisfied  to a great accuracy by the current high
precision data on the magnetic moments \cite{pdb}.

Recently, many efforts have been made to study the baryon magnetic
moments in chiral perturbation theory
\cite{caldi,krause,jenkins,luty,meissner,PP}.  
Clearly, it is very relevant how the $SU(3)$ breaking are treated in such 
study.  In a scheme of chiral perturbation
theory (ChPT)\cite{jenkins} that is popular in the literature the $SU(3)$ 
breaking physical masses are used for strange mesons in the loop.  This 
is the scheme that we shall discuss in this paper.  
(for an alternative scheme, see Ref.\cite{PP}). 
In such scheme, the hyperon magnetic moments receive contributions that are 
non-analytic in the strange quark mass $m_s$, or
equivalently, non-analytic in $SU(3)$ breaking.  Since the Okubo
relation is a result of linear $SU(3)$ breaking, one would
therefore not expect that the moments calculated in ChPT will satisfy
it.  Nevertheless, it was 
shown \cite{caldi} by an explicit calculation
in ChPT that the $\sqrt{m_s}$ corrections to the magnetic moments still
satisfy the Okubo relation.  In more recent applications of ChPT
\cite{jenkins,luty} it was even claimed that both the $\sqrt{m_s}$ and
the $m_s \ln m_s$ corrections satisfy the Okubo relation.  Motivated by
this apparent puzzle, we consider in this paper in more details
the validity of the Okubo relation in this scheme of ChPT.  
We will identify the simple
reason why the $\sqrt{m_s}$ corrections to the hyperon magnetic moments
in this scheme of ChPT satisfy the Okubo relation. 
On the other hand, while our calculational results are in agreement with 
that of Ref.~\cite{jenkins}, our conclusion for the $m_s \ln m_s$ type
non-analytic corrections is different.
We will show that, contrary to the claim in the literature, $m_s \ln
m_s$ non-analytic corrections violate the Okubo relation.

This paper is organized as follows. In the next section we will
introduce the Okubo relation.  In Sec.~\ref{chpt} we will consider the
one-loop diagrams in ChPT that contribute to the magnetic moments, and
will show that they generally do not satisfy the Okubo relation. 
In Sec.~\ref{compare} we will discuss our results and compare them with
earlier calculations.  
In the Appendix, some calculational details are collected.

\section{Okubo relation}

The hyperon magnetic moments satisfy to great accuracy the Okubo
relation \cite{okubo}
\begin{equation}
  \label{OKUBO}
  6\mu_{\Lambda} + \mu_{\Sigma^-} - 4\sqrt{3} \mu_{\Lambda\Sigma^0}
  - 4\mu_n + \mu_{\Sigma^+} - 4 \mu_{\Xi^0} = 0,
\end{equation}
where $\mu_{\Lambda\Sigma^0}$ is the $\Sigma^0\rightarrow\Lambda$
transition moment.  This relation can be obtained \cite{PP} if one
assumes that $SU(3)$ breaking corrections to the moments are linear in
the quark mass matrix $\sigma$, defined by
\begin{equation}
  \label{QMM}
  \sigma \equiv {\rm diag}(0,0,m_s).
\end{equation}
(We neglect the masses of the $u$- and $d$-quark.) In leading order in
the electromagnetic coupling and up to first order in $SU(3)$ symmetry
breaking, the most general expression for the magnetic moments with
this assumption is \cite{krause,PP}
\begin{eqnarray}
\label{MM}
  \mu_{ab} & = & a_1 \langle \tilde{\lambda_b}^\dagger \{ Q ,
  \tilde{\lambda_a} \} \rangle
  + a_2 \langle \tilde{\lambda_b}^\dagger [ Q ,
  \tilde{\lambda_a} ] \rangle
  + \alpha_1 \langle \tilde{\lambda_b}^\dagger [ \sigma ,[ Q , 
  \tilde{\lambda_a} ] ] \rangle
  + \alpha_2 \langle \tilde{\lambda_b}^\dagger \{ \sigma ,[ Q , 
  \tilde{\lambda_a} ] \} \rangle
\nonumber\\&&\mbox{}
  + \alpha_3 \langle \tilde{\lambda_b}^\dagger [ \sigma ,\{ Q , 
  \tilde{\lambda_a} \} ] \rangle
  + \alpha_4 \langle \tilde{\lambda_b}^\dagger \{ \sigma , \{ Q , 
  \tilde{\lambda_a} \} \} \rangle
  + \alpha_5 \langle \tilde{\lambda_b}^\dagger \tilde{\lambda_a}
  \rangle
  \langle \sigma Q \rangle,
\end{eqnarray}
where $a_1$, $a_2$, $\alpha_1, \ldots, \alpha_5$ are arbitrary
parameters, $Q$ is the quark charge matrix
\begin{equation}
  Q = \frac{1}{3} {\rm diag}(2,-1,-1),
\end{equation}
and $\tilde{\lambda_a}$ are generators of $SU(3)$ in the physical
basis, defined e.g.\ in Ref.~\cite{krause,PP}.  Including the transition
moment, \eq{MM} expresses the $9$ magnetic moments in terms of $7$
parameters. Therefore, two relations can be derived between the
moments, which hold for any values of the parameters.  The first is
the Okubo $SU(3)$ relation \eq{OKUBO}, and the second is a simple
isospin relation\cite{PP}.  Motivated by this success, it was argued in 
Ref.\cite{PP} that one should adopt a scheme of ChPT in which the chiral 
symetry breaking appears linearly in $m_s$.  However, a more popular 
scheme of ChPT, as represented in Refs.~\cite{jenkins,luty,meissner}, is 
to use the $SU(3)$ breaking physical meson masses in the loop and as a 
result contributions nonanalytic in strange quark mass are obtained.  The 
puzzle is why Okubo relation is still claimed to be valid within this scheme.

\section{Okubo relation in ChPT}
\label{chpt}

To analyze the Okubo relation for the magnetic moments in this scheme 
of ChPT, we follow the calculations of 
Refs.~\cite{jenkins,luty,meissner}.  The
hyperon magnetic moments receive contributions from the one-meson loop
diagrams in Figs.~(1) and~(2).  Following the scheme in
Refs.~\cite{jenkins,luty,meissner}, we will use an $SU(3)$ invariant mass 
for the baryon propagators inside the loops, while, for the meson 
propagators inside the loops, their physical masses are employed.  All
strong-interaction vertices in the loops originate from the
leading-order $SU(3)$ invariant Lagrangian (see e.g.\ 
Ref.~\cite{jenkins}).  Therefore, the meson propagators and the
electromagnetic vertices are the only sources of $SU(3)$ symmetry
breaking. 

Starting with the loop-diagram in Fig.~(1a), it gives the amplitude
\begin{equation}
\label{amp}
  \Gamma^\mu_{ab} = C_0 \sum_{c,d,e=1}^8 \int {d^4k \over (2\pi)^4} 
  \frac{F_{bce}S_v\cdot (k+q)(G_{ed}(2k+q)^\mu)
  (H_{cad}S_v\cdot k)}{[(k+q)^2-M_e^2 - i\epsilon][k^2-M_d^2
  -i\epsilon][v \cdot k - i\epsilon]},
\end{equation}
where $M_d$ and $M_e$ are the masses of mesons $d$ and $e$,
respectively, $q^\mu$ is the incoming photon four-momentum, and $C_0$ is
a number irrelevant for the following discussion.  The vertex factors
$F_{bce}$, $G_{ed}$, and $H_{cad}$ in \eq{amp} are given by

\begin{equation}
\label{F}
  F_{bce} = D\la\tilde{\lambda}_b^\dagger\{\tilde{\lambda}_e,
  \tilde{\lambda}_c\}\ra
  +F\la\tilde{\lambda}_b^\dagger [\tilde{\lambda}_e,
  \tilde{\lambda}_c]\ra,
\end{equation}
\begin{equation}
  \label{G}
  G_{ed}  = \la\tilde{\lambda}_e^\dagger [ Q,\tilde{\lambda}_d ] \ra,
\end{equation}
and
\begin{equation}
  \label{H}
  H_{cad} = D\la\tilde{\lambda}_c^\dagger\{\tilde{\lambda}_d^\dagger,
  \tilde{\lambda}_a\}\ra
  +F\la\tilde{\lambda}_c^\dagger [\tilde{\lambda}_d^\dagger,
  \tilde{\lambda}_a]\ra,
\end{equation}
respectively.  

The meanings of the flavor indices $a, \ldots, e$ are as illustrated in 
Fig.~(1a).    
Since we took an $SU(3)$ invariant
baryon mass, the flavor index $c$ does not appear in the baryon
propagator. From \eq{G} it is obvious that the electromagnetic vertex
$G_{ed}$ is diagonal in the flavor indices $e$ and $d$, and therefore $M_e = 
M_d$ and the related loop integral depends only on one meson mass. Using this
property it is straightforward to show that the amplitude
$\Gamma^\mu_{ab}$ contributes to the magnetic moment $\mu_{ab}$ as
\begin{equation}
  \label{MUAB1a}
  \mu_{ab}^{(1a)} = 
\sum_{c,e,d=1}^8F_{bce}G_{ed}H_{cad}\;I_{e}, 
\end{equation}
where the functions $I_{e}$ represent the momentum integration part of
the loop diagram.  Assuming isospin symmetry these functions are
readily seen to satisfy 
\begin{equation}
  \label{ident}
  I_{1} = I_{2} = I_{3} = I^{(1a)}(M_\pi^2),\ \
  I_4 = I_5 = I_6 = I_{7} = I^{(1a)}(M_K^2),\ \
  I_8 = I^{(1a)}(M_\eta^2),
\end{equation}
where the functional form of $I$ will be discussed later.

Making similar analysis for the other diagrams in Figs.~(1) and~(2),
one finds that each one-meson loop diagrams will contribute to the
moments $\mu_{ab}$ as
\begin{equation}
\label{MUAB}
  \mu_{ab}(K) = \sum_{e,e'=1}^8 T_{abee'}^K\;I_{e}^K\;\delta_{ee'},
\end{equation}
where, in addition to the meson mass dependence, we add the $K$ index 
to to the integral $I$ to label different diagrams.  For every $K$, the 
functions $I_{e}^K$ obey relations as in \eq{ident}. 
For Fig.~(1a) considered above one obviously has 
\begin{equation}
  \label{EX1}
  T_{abee'}^{(1a)} = \sum_{c,d}^8F_{bce'}G_{ed}H_{cad}.
\end{equation}
For diagrams in Fig.2, each diagram contains only one internal meson line. 
Take Fig.~(2e), which contains an intermediate decuplet propagator, for 
example, one has
\begin{equation}
  \label{EX2}
  T_{abee'}^{(2e)} = \delta_{ed}\;\epsilon_{ijk} 
(\tilde{\lambda}^\dagger_b)^i_l
  (\tilde{\lambda}_{e'})^j_m \Lambda^{klm}
  \bar{\Lambda}_{i'j'k'} (Q)^{k'}_{l'} \Lambda^{i'j'l'}
  \epsilon^{i''j''k''} \bar{\Lambda}_{i''l''m''}
  (\tilde{\lambda}^\dagger_e)^{l''}_{j''}
  (\tilde{\lambda}_a)^{m''}_{k''},
\end{equation}
where $\Lambda_{ijk}=\bar{\Lambda}^{ijk}$ is a basis for the decuplet,
given by \cite{pais}
\begin{eqnarray}
  \Lambda_{ijk} & = & 1,\hspace{1cm}i=j=k,
\nonumber\\
  & = & \frac{1}{3},\hspace{1cm}i=j\ne k,
\nonumber\\
  & = & \frac{1}{6}, \hspace{1cm} i,j,k\ {\rm distinct}.
\end{eqnarray}
The internal octet and decuplet indices in $T_{abee'}$ can be easily
summed over using
\begin{equation}
  \label{fierz}
  \sum_{a=1}^8(\tilde{\lambda}_a^\dagger)^\alpha_\beta
  (\tilde{\lambda}_a)^\gamma
  _\rho = 2\delta^\alpha_\rho\delta^\gamma_\beta
  -{2\over 3}\delta^\alpha_\beta
\delta^\gamma_\rho,
\end{equation}
and
\begin{equation}
  \label{DECUPLET}
  \Lambda^{klm}\bar{\Lambda}_{k'l'm'} = \frac{1}{6}(
  \delta^k_{k'}\delta^l_{l'} \delta^m_{m'} +
  \delta^k_{k'}\delta^m_{l'}\delta^l_{m'} +
  \delta^l_{k'}\delta^k_{l'}\delta^m_{m'} +
  \delta^l_{k'}\delta^m_{l'}\delta^k_{m'} +
  \delta^m_{k'}\delta^k_{l'}\delta^l_{m'} +
  \delta^m_{k'}\delta^l_{l'}\delta^k_{m'}). 
\end{equation} 
The general result after these summations is that $T_{abee'}$ in
\eq{MUAB} is a flavor structure constructed out the remaining matrices
$\tilde{\lambda}_a$, $\tilde{\lambda}_b^\dagger$, $\tilde{\lambda}_{e'}$,
$\tilde{\lambda}_e^\dagger$, and $Q$, in which each of these matrices
appears exactly one time.  In fact, $T_{abee'}$ would be a $SU(3)$ tensor 
if not for the $SU(3)$ breaking matrix $Q$.  The fact that all the one-meson 
loops in Figs.~(1) and~(2) can be written in the form of \eq{MUAB} makes it
possible to examine their $SU(3)$ symmetry breaking pattern in a simple way.

To analyze the $SU(3)$ symmetry structure of the magnetic moments in 
\eq{MUAB}, 
we first note that, for each $K$, $I_{e}^K$ can be rewritten as 
\begin{equation}
  \label{I}
  I_{e}\delta_{ee'}  = \langle \tilde{\lambda}^\dagger_{e'} \{
  \hat{I} , \tilde{\lambda}_e \} \rangle +
  \Bigl[ I(M_\eta^2)-{4\over 3}I(M_K^2)+{1\over 3}I(M_\pi^2) \Bigr]
  \delta_{e'8}\delta_{e8},
\end{equation}
where $\hat{I}$ is the following $3\times 3$ matrix
\begin{equation}
\label{hatI}
  \hat{I} = \frac{1}{2} I(M_\pi^2)
  +  \frac{1}{m_s}[I(M_K^2)-I(M_\pi^2)]\sigma,
\end{equation}
which is a simple linear combination of the unit matrix and the quark
mass matrix.  Note that each integration factor, $I$, in 
in Eqs.(\ref{I}) and (\ref{hatI}) have diagramatic, $K$, dependence as 
well which we supressed in notation.  
Using \eq{I} one can then split the magnetic moments $\mu_{ab}$ in
two parts as
\begin{equation}
  \mu_{ab} = \mu_{ab}^{\rm I} + \mu_{ab}^{\rm II},
\label{MU}
\end{equation}
where  $\mu_{ab}^{\rm I}$ is defined by
\begin{equation}
  \label{MU1}
  \mu_{ab}^{\rm I} = \sum_{K}\sum_{e,e'=1}^8 T_{abee'}^K
  \langle \tilde{\lambda}^\dagger_{e'} \{
  \hat{I}^K , \tilde{\lambda}_e \} \rangle, 
\end{equation}
and $\mu_{ab}^{\rm II}$ is defined by
\begin{equation}
  \label{MUII}
  \mu_{ab}^{\rm II} = 
  \sum_{K} \mu_{ab}^{\rm II}(K) = 
\sum_{K}\Bigl[ (I^K(M_\eta^2)-{4\over 3}I^K(M_K^2)+
  {1\over 3}I^K(M_\pi^2) \Bigr] T_{ab88}^K.
\end{equation}
Note that because the $\eta$ meson is charge neutral, the electromagnetic 
vertex $G_{ed}$ is zero for $\eta$ loop in Figs.(1a,b).  As a result, 
Figs.(1a,b) contribute only to $\mu_{ab}^{\rm I}$ because 
$T_{ab88}^{(1a,b)}=0$.  Figs.~(2) can contribute to both $\mu_{ab}^{\rm 
I,II}$.

The symmetry properties of $\mu_{ab}^{\rm I}$ can be easily analyzed.
For the summations over $e$ and $e'$ in \eq{MU1} one can again use the
completeness relation \eq{fierz}.  Since the matrix $\hat{I}$ is a
linear combination of the unit matrix and the quark mass matrix
$\sigma$, and given the properties of $T_{abee'}$ as discussed after
\eq{DECUPLET}, one finds that $\mu_{ab}^{\rm I}$ is a tensor formed
out of the matrices $\tilde{\lambda}_a$, $\tilde{\lambda}_b^\dagger$,
$Q$, and $\sigma$, in which $\tilde{\lambda}_a$,
$\tilde{\lambda}_b^\dagger$, and $Q$ each appears exactly one time, and
the matrix $\sigma$ appears at most one time. The general form of such
a tensor is given by \eq{MM}.  As was discussed after \eq{MM}, the
$\mu_{ab}^{\rm I}$ component of the magnetic moments 
therefore satisfy the Okubo relation \eq{OKUBO} no matter what kind of 
meson mass dependences result from the integration part of the contribution.

Whether or not the magnetic moments $\mu_{ab}$, \eq{MUAB}, satisfy the
Okubo relation then depends on whether or not the moments
$\mu_{ab}^{\rm II}$, \eq{MUII}, satisfy the Okubo relation.  
As a result we only have to worry about the contributions of diagrams in 
Figs(2).  For any
given one-loop diagram, the contribution to the magnetic moment from
only the $\eta$-loop, denoted here by $\mu_{ab}^\eta$, is, according to
\eqs{MUAB}{ident}, given by
\begin{equation}
  \label{ETA}
  \mu_{ab}^\eta(K) = I^K(M_\eta^2) T_{ab88}^K.
\end{equation}
Using \eq{ETA} we can relate $\mu_{ab}^{\rm II}$ to the $\eta$-loop as
\begin{equation}
  \label{MU2}
  \mu_{ab}^{\rm II} = 
\sum_{K}\frac{\Delta^K}{3I^K(M_\eta^2)}\mu_{ab}^\eta(K), \end{equation}
where we have defined
\begin{equation}
  \label{DELTA}
  \Delta^K = 3 I^K(M_\eta^2)-4 I^K(M_K^2)+I^K(M_\pi^2). 
\end{equation}
Since the meson masses are non-degenerate, $\Delta^K$ in \eq{MU2} is in
general nonzero.  The magnitude of $\Delta^K$ depends on the function
$I^K$ in \eq{DELTA}.  By closer inspection one finds that the
generic form of $I$ for the diagrams in Fig.~(1) is given by
\begin{equation}
  \label{FIG1}
  I^K(X^2) = A_1^K + A_2^K \sqrt{X^2},
\end{equation}
while the generic form of $I^K$ for the diagrams in Fig.~(2) is given by
\begin{equation}
  \label{FIG2}
  I^K(X^2) = B_1^K + B_2^K X^2 \ln \left( X^2 / \mu^2 \right) , 
\end{equation}
where the coefficients $A_1^K$, $A_2^K$, $B_1^K$, and $B_2^K$ depend 
on the diagrams but are independent of 
$X$, and $\mu$ is the renormalization scale.
Note that in terms of the quark masses $m_q$, \eq{FIG1} leads to
corrections of the form $\sqrt{m_q}$, and \eq{FIG2} leads to
corrections of the form $m_q \ln m_q$.  We conclude immediately here that 
the nonanalytic contributions of the form $\sqrt{m_q}$ satisfy the Okubo 
relation.  This part, we agree with Refs.~\cite{jenkins,luty}.

By substituting the functional forms \eq{FIG2} into \eq{DELTA}, the
diagrams in Figs.~(12) all yield $\Delta^K \ne 0$.  Therefore, by
\eq{MU2}, $\mu_{ab}^{\rm II}$ is proportional to the $\eta$-loop
contribution to the magnetic moments. For the one-loop diagrams
contributing to the hyperon magnetic moments we then conclude: A
diagram satisfies the Okubo relation if and only if its $\eta$ loop 
contribution, or equivalently its $T_{ab88}$ term, satisfies the Okubo 
relation.  This observation greatly simplifies the
analysis of the validity of the Okubo relation of hyperon magnetic
moments calculated in current scheme of ChPT.

Finally we note that $\mu_{ab}^{\rm II}$ is explicitly renormalization 
scale independent. 
This can be seen by substituting \eq{FIG2} in \eq{DELTA}, we find that
\begin{equation}
\label{REN}
\Delta^K = - B_2(K) \left[ 3 M_\eta^2 - 4 M_K^2 + M_\pi^2 \right] \ln 
	\mu^2 
	 + ({\it renormalization\;\; independent\;\; terms} ) , 
\end{equation}
The dependence on $\mu$ in \eq{REN} vanishes upon using the 
Gell-Mann-Okubo relation for the meson masses.
On the other hand, $\mu_{ab}^{\rm I}$ is renormalization scale dependent, 
however, since they are of the form as in Eq.(3), they can be removed by 
introducing renormalization counterterms.  We can now discuss the 
consequence of the above analysis.


\section{Discussions and summary}
\label{compare}

>From the discussion in the previous
section it follows that contributions from diagrams
which have a vanishing $\eta$-loop, will trivially satisfy the Okubo
relation.  This is the case for the diagrams in Fig.~(1).  
This gives a simple explanation why the contributions of the diagrams in
Fig.~(1), which give rise to the $\sqrt{m_q}$ corrections, satisfy the 
Okubo relation consistent with the observations in 
Refs.~\cite{caldi,jenkins}. 

For the same reason, the
diagram Fig.~(2b,2c) yields magnetic moments that satisfy the Okubo
relation.  For Fig.~(2b), it is because the $\eta$ meson is charge 
neutral.  For Fig.~(2c), it is purely due to accidental cancellation as 
shown in the Appendix.

On the other hand, as shown explicitly in the Appendix, the $\eta$ 
loop contribution to the magnetic moments from each of the diagrams 
Figs.~(2a), (2d), (2e), (2f), (2g), (2h) and (2i) is non-vanishing and 
violates 
the Okubo relation.  They all give contributions to the hyperon moments 
of the form $m_q \ln m_q$.  We also show that their sum also remains nonzero.
While our detail analytic result, as demonstrated in the Appendix, agrees 
with Refs.~\cite{jenkins}, our conclusion is in conflict with that in 
Refs.~\cite{jenkins,luty} which claimed that corrections of the form $m_q 
\ln m_q$ satisfy the Okubo relation.  The magnitude of the
deviation from the Okubo relation will in general depend on the
parameters from the ChPT Lagrangian and the meson masses.  There is 
no reason, based upon $SU(3)$ symmetry arguments alone, a priori that
this deviation should vanish or should be small.  

One should finally also note that in the modified scheme of $SU(3)$ ChPT 
proposed in Ref.~\cite{PP,PPP}, loop diagrams give rise to linear $SU(3)$
symmetry breaking automatically in the lowest nontrivial order in 
$SU(3)$ breaking.  In that case one can easily find that $\Delta^K=0$ 
automatically in \eq{MU2}, and the magnetic moments satisfy the Okubo 
symmetry relation.

The hyperon magnetic moments satisfy to a high precision the Okubo
relation.  This relation is based on the assumption that $SU(3)$ is
broken linearly in the strange quark mass $m_s$.  In this paper we have
studied the validity of the Okubo relation if the moments are
calculated in a scheme of $SU(3)$ chiral perturbation theory (ChPT) at the 
one-loop level.

We have found a simple way to classify the one-loop diagrams of this ChPT
contributing to the magnetic moments according to whether or not they
generate magnetic moments that satisfy the Okubo $SU(3)$ relation. A
diagram satisfies the Okubo relation if and only if its $\eta$ loop
satisfies the Okubo relation. Using this observation, we have shown
that the $\sqrt{m_s}$ corrections satisfy the Okubo relation. However,
some of the $m_s \ln m_s$ corrections violate the Okubo relations as
opposed to what was claimed in recent literature.  Note that the above
analysis could in principle be extended to the scheme in which that $SU(3)$
breaking masses are allowed even in the baryon propagators.

\section*{Acknowledgments}

This work was partially supported by grants from the National Science 
Council of the Republic of China.  DC also wish to thank the NSC Center 
for Theoretical Sciences for partial support.

\section*{Appendix}

In this appendix we give the deviation from the Okubo relation among the 
hyperon magnetic moments for each one-loop diagrams in Figs. (1)
and (2) and the
total deviation.

According to the discussion in Sec.\ref{chpt}, the deviation from the Okubo 
relation for each one-loop diagram only depends on the second term 
$\mu_{ab}^{\rm II}$ in Eq.(\ref{MU}) and can be expressed as
\begin{eqnarray}
\label{devi_oku}
 \Omega(K) &=& 6\mu_{\Lambda}^{\rm II}(K) + \mu_{\Sigma^-}^{\rm II}(K) 
	- 4\sqrt{3} \mu_{\Lambda \Sigma}^{\rm II}(K) - 4\mu_n^{\rm II}(K) 
\nonumber \\
	   & & + \mu_{\Sigma^+}^{\rm II}(K) - 4 \mu_{\Xi^0}^{\rm II}(K),
\end{eqnarray}
where $K$ label different diagrams. Using Eqs.(\ref{MUII}) and 
(\ref{DELTA}), the 
$\Omega(K)$ can be denoted as
\begin{equation}
\Omega(K) = \frac{1}{3}\; \Delta^K\; T(K)
\end{equation}
where $\Delta(K)$ is defined in Eq.(\ref{DELTA}) and
\be
T(K) &=& 6 T_{8888}^K+T_{2288}^K-4\sqrt{3}T_{8388}^K-4T_{6688}^K
\nonumber \\
     & & +T_{1188}^K-4T_{7788}^K.
\en

The $T_{ab88}$ is expected to be nonzero only for the diagonal components in 
the flavor indices (a,b) i.e. a=b, and  for (a,b)=(3,8) or (8,3) 
, which is related to the transition magnetic moment $\mu_{\Lambda\Sigma}$.
In the following, we give the $T_{ab88}$ for each diagram, and, at the 
second line of each equation, we list explicitly only the values of diagonal 
components in indices (a,b) and component (a,b)=(8,3) in the order 
(a,b)=\{(1,1),(2,2),(3,3),(4,4),(5,5),(6,6),(7,7),(8,8),(8,3)\}, 
corresponding to the magnetic moments in the order 
($\mu_{\Sigma^+}$,$\mu_{\Sigma^-}$,$\mu_{\Sigma^0}$,$\mu_p$,$\mu_{\Xi^-}$,
$\mu_n$,$\mu_{\Xi^0}$, \\
$\mu_\Lambda$,$\mu_{\Lambda\Sigma}$).
\be
  \label{EX1a}
  T^{(1a)}_{ab88} &=&\sum_{c}^8F_{bc8}G_{88}H_{ca8},
\nonumber \\
	&=& \{0,0,0,0,0,0,0,0,0\}
\en
\be
  \label{EX1b}
  T^{(1b)}_{ab88} &=& \sum_{c}^8T_{bc8}G_{88}R_{ca8},
\nonumber \\
	&=& \{0,0,0,0,0,0,0,0,0\}
\en
\be 
\label{EX2a}
  T^{(2a)}_{ab88} &=& \sum_{c,d}^8F_{bc8}V_{cd}H_{da8},
\nonumber \\
	&=& \{\frac{2}{9}D^2(\mu_D+3\mu_F), \frac{2}{9}D^2(\mu_D-3\mu_F),
	     \frac{2}{9}D^2\mu_D, 
\nonumber \\
	& & \frac{1}{18}(D-3F)^2(\mu_D+3\mu_F), 
	     \frac{1}{18}(D+3F)^2(\mu_D-3\mu_F), 
\nonumber \\
	& & -\frac{1}{9}(D-3F)^2\mu_D, -\frac{1}{9}(D+3F)^2\mu_D, 
	     -\frac{2}{9}D^2\mu_D,
\nonumber \\
	& &  -\frac{2}{3\sqrt{3}}D^2\mu_D\}
\en
\be
\label{EX2b}
  T^{(2b)}_{ab88} &=& \la\tilde{\lambda}_b^\dagger [ 
[\tilde{\lambda_8},
        , [ \tilde{\lambda}_8^\dagger, Q]] ,\tilde{\lambda}_a ] \ra,
\nonumber \\
  	&=& \{0, 0, 0, 0, 0, 0, 0, 0, 0\}
\en
\be
  \label{EX2c}
  T^{(2c)}_{ab88} &=& \sum_{c,d}^8T_{bc8}U_{cd}R_{da8},
\nonumber \\
	&=& \{ 6{\cal C}^2\mu_C, -6{\cal C}^2\mu_C, 0, 6{\cal C}^2\mu_C,
	       0, 0, 0, 0, 0, 0 \}
\en
\be
  \label{EX2d}
  T^{(2d)}_{ab88} &=& \sum_{c,d}^8T_{bc8}X_{cd}H_{da8},
\nonumber \\
	&=& \{-2D{\cal C}\mu_T, 0, -D{\cal C}\mu_T, 0, 0, 0, 
	     (D+3F){\cal C}\mu_T, 
\nonumber \\
	& & 0, -\sqrt{3}D{\cal C}\mu_T\}
\en
\be
  \label{EX2e}
  T^{(2e)}_{ab88} &=& \sum_{c,d}^8T_{bc8}X_{cd}H_{da8},
\nonumber \\
	&=& \{-2D{\cal C}\mu_T, 0, -D{\cal C}\mu_T, 0, 0, 0, 
	     (D+3F){\cal C}\mu_T, 
\nonumber \\
	& & 0, 0\}
\en
\be
  \label{EX2f}
  T^{(2f)}_{ab88} &=& \sum_{c}^8F_{bc8}H_{cb8}V_{ba},
\nonumber \\
	&=& \{\frac{2}{9}D^2(\mu_D+3\mu_F), \frac{2}{9}D^2(\mu_D-3\mu_F),
	     \frac{2}{9}D^2\mu_D, 
\nonumber \\
	& & \frac{1}{18}(D-3F)^2(\mu_D+3\mu_F), 
	     \frac{1}{18}(D+3F)^2(\mu_D-3\mu_F), 
\nonumber \\
	& & -\frac{1}{9}(D-3F)^2\mu_D, -\frac{1}{9}(D+3F)^2\mu_D, 
	     -\frac{2}{9}D^2\mu_D,
\nonumber \\
	& & \frac{2}{3\sqrt{3}}D^2\mu_D\}
\en
\be
  \label{EX2g}
  T^{(2g)}_{ab88} &=& \sum_{c}^8V_{ba}F_{ac8}H_{ca8},
\nonumber \\
	&=& \{\frac{2}{9}D^2(\mu_D+3\mu_F), \frac{2}{9}D^2(\mu_D-3\mu_F),
	     \frac{2}{9}D^2\mu_D, 
\nonumber \\
	& & \frac{1}{18}(D-3F)^2(\mu_D+3\mu_F), 
	     \frac{1}{18}(D+3F)^2(\mu_D-3\mu_F), 
\nonumber \\
	& & -\frac{1}{9}(D-3F)^2\mu_D, -\frac{1}{9}(D+3F)^2\mu_D, 
	     -\frac{2}{9}D^2\mu_D,
\nonumber \\
	& & \frac{2}{3\sqrt{3}}D^2\mu_D\}
\en
\be
  \label{EX2h}
  T^{(2h)}_{ab88} &=& \sum_{c}^8T_{bc8}R_{cb8}V_{ba},
\nonumber \\
	&=& \{ 2{\cal C}^2(\mu_D+3\mu_F), 2{\cal C}^2(\mu_D-3\mu_F),
	      2{\cal C}^2\mu_D, 
\nonumber \\
	& &  0, 2{\cal C}^2(\mu_D-3\mu_F), 0, -4{\cal C}^2\mu_D,  0, 0\}
\en
\be
  \label{EX2i}
  T^{(2i)}_{ab88} &=& \sum_{c}^8V_{ba}T_{ac8}R_{ca8},
\nonumber \\
	&=& \{ 2{\cal C}^2(\mu_D+3\mu_F), 2{\cal C}^2(\mu_D-3\mu_F),
	      2{\cal C}^2\mu_D, 
\nonumber \\
	& &  0, 2{\cal C}^2(\mu_D-3\mu_F), 0, -4{\cal C}^2\mu_D,  0, 
	    2\sqrt{3}{\cal C}^2\mu_D\}
\en

where
\begin{equation}
\label{T}
T_{bac} = {\cal C} \epsilon_{ijk}(\tilde{\lambda}_b^\dagger)_l^i 
	  (\tilde{\lambda}_c)_m^j\Lambda^{klm}
\end{equation}
\begin{equation}
\label{R}
R_{bac} = {\cal C} \epsilon^{ijk}\bar{\Lambda}_{klm} 
	  (\tilde{\lambda}_c^\dagger)^m_j(\tilde{\lambda}_a)_i^l
\end{equation}
\begin{equation}
\label{V}
V_{ba} = \mu_D\la\tilde{\lambda}_b^\dagger \{ Q, \tilde{\lambda}_a \} \ra
	+ \mu_F\la\tilde{\lambda}_b^\dagger [ Q, \tilde{\lambda}_a ] \ra,
\end{equation}
\begin{equation}
\label{U}
U_{ba} = \mu_C \bar{\Lambda}_{ijk}( Q )_l^k \Lambda^{ijl}
\end{equation}
\begin{equation}
\label{X}
X_{ba} = \mu_T \epsilon^{ijk} Q_i^l (\bar{\Lambda})_{klm}
	 (\tilde{\lambda}_a)_j^m
\end{equation}
\begin{equation}
\label{Y}
Y_{ba} = \mu_T \epsilon_{ijk} Q_l^i (\tilde{\lambda}_b^\dagger)_m^j
	 \Lambda^{klm}
\end{equation}
with the coupling constants $D$, $F$, ${\cal C}$, $\mu_D$, $\mu_F$, $\mu_C$, 
and $\mu_T$ defined in Ref.\cite{jenkins}.
The vertex factors $F_{bac}$, $G_{ba}$, and $H_{bac}$ are given in Eqs. 
(\ref{F}), (\ref{G}), and (\ref{H}), respectively.

\begin{table}[t]
\caption{The deviation $T(K)$ from the Okubo relation for the 
octet baryon magnetic moments in each label $K$ diagram. All moments are 
given in nuclear magnetons}
\label{table1}
\centerline{
\begin{tabular}{||c|c||} \hline
   $K$   & $T(K)$ \\
\hline
Fig.(1a) & 0 \\
\hline
Fig.(1b) & 0 \\
\hline
Fig.(2a) & $\frac{8}{3} \mu_D ( D^2 + 3 F^2 )$ \\
\hline
Fig.(2b) & 0 \\
\hline
Fig.(2c) & 0 \\
\hline
Fig.(2d) & $ \mu_T {\cal C} ( D - 2 F )$ \\
\hline
Fig.(2e) & $- \mu_T {\cal C} ( D + 2 F )$ \\
\hline
Fig.(2f) & $-\frac{8}{3} \mu_D ( D^2 - 3 F^2 )$ \\
\hline 
Fig.(2g) & $-\frac{8}{3} \mu_D ( D^2 - 3 F^2 )$ \\
\hline
Fig.(2h) & $ \frac{5}{3} \mu_D {\cal C}^2$ \\
\hline
Fig.(2i) & $-\frac{1}{3} \mu_D {\cal C}^2$ \\
\hline
\end{tabular} 
}
\end{table}

All the one-loop diagrams in Fig. (2) contribute the $m_q \ln m_q$
corrections to the hyperon magnetic moments, but only the 
contributions of the diagrams in 
Figs. (2b) and (2c) explicitly satisfy the Okubo relation.

For the diagrams in Figs.(2a), (2d), (2e), (2f), (2g), (2h) and (2i) 
the fact that $T(K) \neq 0$ individually means that each diagram 
violates the Okubo relation.
However, one must check explicitly whether or not there is any 
cancellation when these contributions from different diagrams add up.
According to their coupling constant factors,
the Okubo relation violating one-loop diagrams can be classified into 
three groups: (I) diagrams (2a), (2f), (2g) with factor 
$\mu_D D^2$ or $\mu_D F^2$; (II) diagrams (2d), (2e) with factor 
$\mu_T{\cal C}D$ or $\mu_T{\cal C}F$; and (III) diagrams (2h), (2i) with 
factor $\mu_D \cdot {\cal C}^2$. 

The Feynman integral $I$ in Eq. (\ref{MUII}) for diagram Fig. (2a) is
\be
\label{I2a}
I^{(2a)}(M^2) 
	    =  -\frac{1}{32\pi^2f^2} M^2 \ln(\frac{M^2}{\mu^2})   \;,
\en
where $\mu$ is an arbitrary renormalization scale 
and $M$ is the physical mass of the intermediate meson state.
The contributions from Figs. (2f) and (2g) correspond to wave function 
renormalization with intermediate octet baryon. The integrals I for both 
diagrams turn out to be the same and equal to
\be
\label{I2f}
I^{(2f)}(M^2) &=& -\frac{3}{2}\cdot\frac{1}{32\pi^2f^2} 
	M^2 \ln(\frac{M^2}{\mu^2}) \;.
\en

Summing over the contributions from Figs. (2a), (2f), and (2g), we get 
\be
& & \sum_{K=Fig. (2a),(2f),(2g)}\Omega(K) 
\nonumber \\
&=& \frac{1}{32\pi^2f^2} [ 3M_\eta^2 \ln(\frac{M_\eta^2}{\mu^2})
	- 4M_K^2 \ln(\frac{M_K^2}{\mu^2})+M_\pi^2 \ln(\frac{M_\pi^2}{\mu^2})]
	[\frac{16}{9}\mu_D (D^2-6F^2)]  \;.
\nonumber \\
\en

For the diagrams in Figs. (2d) and (2e), using the SU(3) invariant mass 
values for intermediate octet and decuplet 
baryons, the one-loop integrals $I(M^2)$ for both diagrams
can be easily shown to be equal and 
\be
\label{I2d}
I^{(2d)}(M^2) &=& \frac{2}{3}\cdot\frac{1}{32\pi^2f^2} M^2 
\ln(\frac{M^2}
	{\mu^2})   \;.
\en

Figs. (2h) and (2i) are related to the wave function renormalization with
the intermediate decuplet baryon. 
The integration function $I(M^2)$ for Fig. (2h) or (2i) is
\be
\label{I2h}
I^{(2h)}(M^2) &=&  -\frac{1}{32\pi^2f^2} M^2 
\ln(\frac{M^2}{\mu^2})   \;. \en

Summing over all the contributions in 
Fig.(2), the deviation from the Okubo relation is
\be
\Omega 	&=& \sum_{K}\Omega(K)
\nonumber \\
&=& \frac{1}{32\pi^2f^2} [ 3M_\eta^2 \ln(\frac{M_\eta^2}{\mu^2})
	- 4M_K^2 \ln(\frac{M_K^2}{\mu^2})+M_\pi^2 \ln(\frac{M_\pi^2}{\mu^2})]
\nonumber \\
& &	[\frac{4}{9}\mu_D (4D^2-24F^2-{\cal C}^2)+\frac{8}{9}\mu_T {\cal C}F]  
\;.
\en
If we add up the contributions listed in Ref.\cite{jenkins} from various 
diagrams, the result actually consistent with our sum above.
However, unless there is a very special relation among $D$, $F$, ${\cal C}$, 
$\mu_D$ and $\mu_T$, which is not the case, the Okubo relation is indeed 
violated by the logarithmic corrections in $m_q$ contrary to the claim in 
Ref.\cite{jenkins}.

\newpage
\section*{Figure captions}

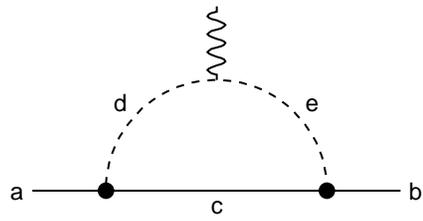
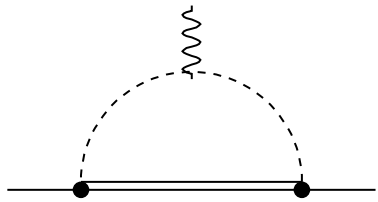
{\bf Fig. 1:} One-loop \noindent diagrams that give rise to non-analytic
$\sqrt{m_q}$ corrections to the hyperon magnetic moments.
The dashed lines denote the pions, the single solid lines
denote octet hyperons, and the double solid lines denote decuplet hyperons
\vspace{0.5cm}
\\
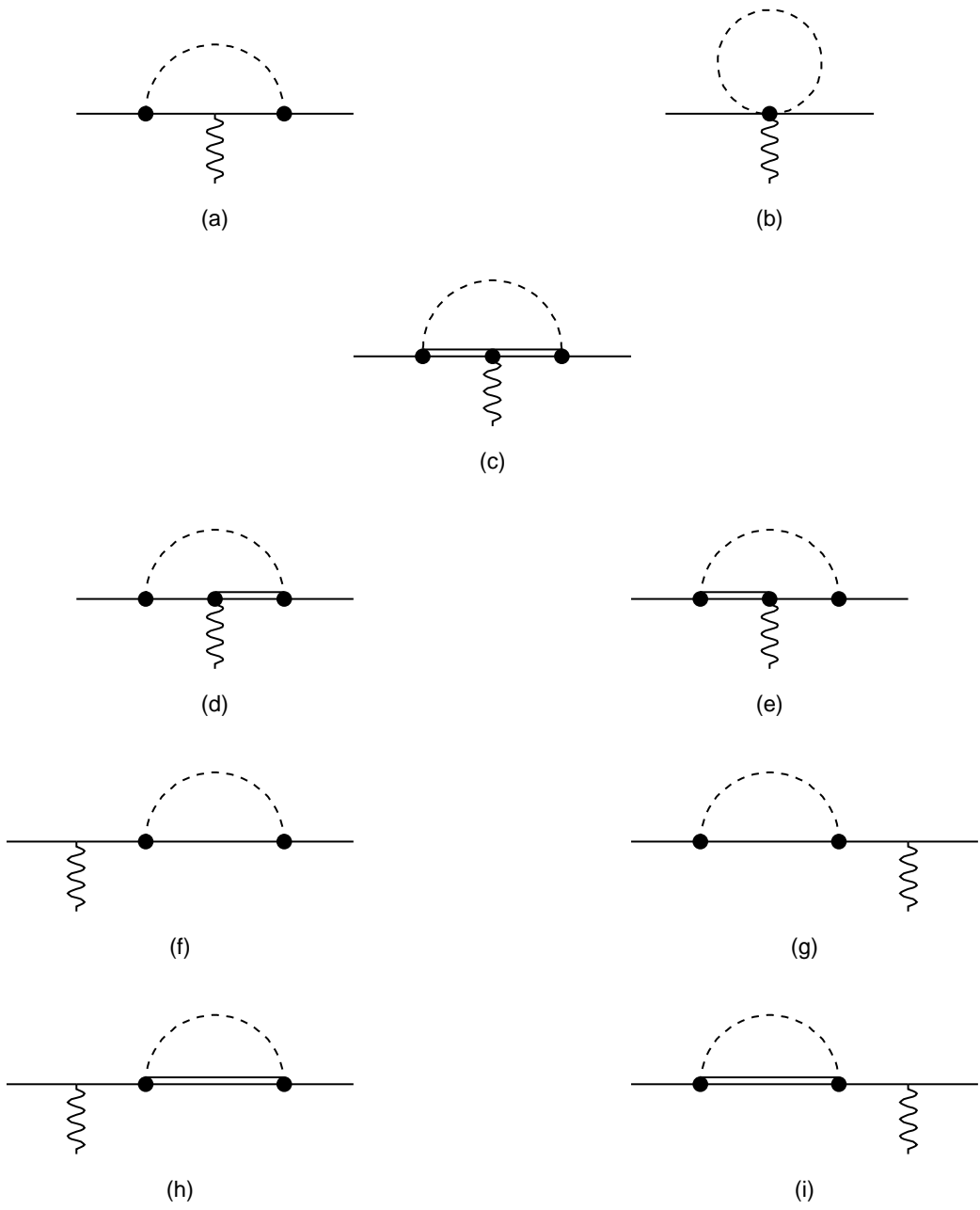
{\bf Fig. 2:} One-loop diagrams that give rise to non-analytic
$m_q \ln m_q$ corrections to the hyperon magnetic moments.
See Fig.~(1) for the meaning of the lines.

\end{document}